\documentclass{PoS}

\usepackage{graphicx}
\usepackage{amsmath}
\usepackage{amssymb}
\usepackage{slashed}

\newcommand{\tr}{\rm tr \,}

\newcommand{\CD}{\hat \partial}

\title{On chiral extrapolations of coupled-channel reaction dynamics for charmed mesons}

\ShortTitle{On chiral extrapolations of coupled-channel reaction dynamics for charmed mesons}

\author{\speaker{Xiao-Yu Guo}\\
        GSI Helmholtzzentrum f\"ur Schwerionenforschung GmbH, \\Planckstra\ss e 1, 64291 Darmstadt, Germany\\
        E-mail: \email{x.guo@gsi.de}}
\author{Yonggoo Heo\\
        Suranaree University of Technology, Nakhon Ratchasima, 30000, Thailand\\
        E-mail: \email{y.heo@g.sut.ac.th}}  
\author{Matthias F.M. Lutz\\
        GSI Helmholtzzentrum f\"ur Schwerionenforschung GmbH, \\Planckstra\ss e 1, 64291 Darmstadt, Germany\\
        Technische Universit\"at Darmstadt, D-64289 Darmstadt, Germany\\
        E-mail: \email{m.lutz@gsi.de}}       

\abstract{We perform an analysis of QCD lattice data on masses of charmed meson with $J^P = 0^-$ and $J^P =1^-$ quantum numbers. The quark-mass dependence of the data set is used to gain information on the size of counter terms in the chiral Lagrangian formulated with open-charm mesons. Of particular interest are those counter terms that are active in the exotic flavour sextet channel. A chiral expansion scheme is developed and applied to the lattice data set. An accurate reproduction of the lattice data based on ensembles 
of PACS-CS, MILC, ETMC and  HSC with pion and kaon masses smaller than 600 MeV is achieved.
It is argued that a unique set of low-energy parameters is obtainable only if additional information from an HSC ensemble on some scattering phase shifts is included in our global fits. Based on such low-energy parameters we find a clear signal for a member of the exotic flavour sextet states in the $\pi D$ phase shift, between the $\eta D$ and 
$\bar K D_s$ thresholds. A striking dependence of such phase shifts on the values of the up, down and strange quarks is predicted.  }

\FullConference{The 36th Annual International Symposium on Lattice Field Theory - LATTICE2018\\
		22-28 July, 2018\\
		Michigan State University, East Lansing, Michigan, USA.}

\begin{document}

\section{Introduction}

Open-charm meson systems are believed to be of crucial importance in understanding the low-energy behavior of QCD \cite{Yan:1992gz,Lutz:2015ejy}. A heavy charm quark is surrounded by a light quark, either a  $u, d$ or $s$ quark. The interplay of the heavy-quark spin symmetry for the charm quark and the chiral  SU(3) symmetry for the $u, d$ or $s$ quarks make such systems unique  and  effective field theory approaches particularly predictive. The leading order chiral Lagrangian already produces significant short-range and attractive forces that may dynamically generate open-charm resonances. Within the coupled-channel approach \cite{Lutz:2001yb}, the scalar meson $D_{s0}^*(2317)$ is well described by the leading order chiral interaction \cite{Kolomeitsev:2003ac,Hofmann:2003je,Lutz:2007sk,Guo:2006fu,Guo:2015dha,Du:2017ttu}. At this leading order further predictions are made on exotic scalar resonances in a flavour sextet \cite{Kolomeitsev:2003ac,Hofmann:2003je,Lutz:2007sk}. 
The existence of such states depends on the precise form of chiral forces generated by higher order chiral counter terms \cite{Hofmann:2003je,Lutz:2007sk,Liu:2012zya,Lutz:2015ejy}.
The aim of this work is to report on the impact of lattice data from \cite{Mohler:2011ke,Na:2012iu,Kalinowski:2015bwa,Moir:2016srx} on such counter terms and their implications for coupled-channel open-charm systems.

We notice that the counter terms have significant impact not only on the open-charm coupled channel systems but also on the quark-mass dependence of the $D$-meson masses. The lattice data sets on charmed meson masses from \cite{Mohler:2011ke,Na:2012iu,Kalinowski:2015bwa} are considered and play a crucial role in our derivation of a set of the low-energy parameters.
Such data are supplemented by first lattice data from HSC on the  s-wave $\pi D$ scattering process at a pion mass of about 400 MeV \cite{Moir:2016srx}. Based on our fit scenarios, we present the pole positions found for the scalar open-charm mesons in  the exotic flavour sextet channels as expected for the physical choice of the quark masses. A striking quark-mass dependence of such states is illustrated by presenting the form of the s-wave $\pi D$ phase shift as extrapolated down from the considered HSC ensemble at a pion mass of about 400 MeV to the physical point.

\section{The chiral Lagrangian for open-charm mesons}
\label{sec:2}

We recall the SU(3) chiral Lagrangian formulated in the presence of anti-triplets of $D$ mesons with $J^P =0^-$ and $J^P =1^-$. In the relativistic version the Lagrangian was developed in \cite{Kolomeitsev:2003ac,Hofmann:2003je,Lutz:2007sk}.  The chiral Lagrangian used in \cite{Guo:2018kno} considers 
the anti-triplets fields, $D$ and $D_{\mu \nu}$, of charmed mesons with $J^P= 0^-$ and $J = 1^-$ quantum numbers respectively. The terms relevant here read
\begin{eqnarray}
&& \mathcal{L}_{}=(\CD_\mu D)(\CD^\mu \bar D)-  M^2\, D \, \bar D
  + 2\,g_P\,\big\{D_{\mu \nu}\,U^\mu\,(\CD^\nu \bar D)
 - (\CD^\nu D )\,U^\mu\,\bar D_{\mu \nu} \big\}
\nonumber\\ 
&& \quad - \,\big( 4\,c_0-2\,c_1\big)\, D \,\bar{D}  \,{\tr} \chi_+ -2\,c_1\,D \,\chi _+\,\bar{D}
 + \, 4\,\big(2\,c_2+c_3\big)\,D\bar{D}\,{\tr} \big(U_{\mu }\,U^{\mu \dagger }\big)- 4\,c_3\, D \,U_{\mu }\,U^{\mu \dagger }\,\bar{D}
\nonumber\\
&& \quad +\,\big( 4\,c_4+2\,c_5\big)\, ({\CD_\mu } D) ({\CD_\nu }\bar{D}) \,{\tr} \big[ U^{\mu }, \,U^{\nu \dagger }\big]_+ /M^2
-2\,c_5\,({\CD_\mu } D) \big[ U^{\mu }, \,U^{\nu \dagger }\big]_+({\CD_\nu }\bar{D}) /M^2
\nonumber\\
&& \quad + \, 4\,g_1\,{D}\,[\chi_-,\,{U}_\nu]_- \CD^\nu \,\bar{D}/M
  - 4\,g_2\,{D}\,\big([{U}_\mu,\,[\CD_\nu,\,{U}^\mu]_-]_- + [{U}_\mu,\,[\CD^\mu,\,{U}_\nu]_-]_-\big)\,\CD^\nu\bar{D}/M
\nonumber\\ 
&& \quad  - \,4\,g_3\,{D}\,[{U}_\mu,\,[\CD_\nu,\,{U}_\rho]_-]_-\,[\CD^\mu,\,[\CD^\nu,\,\CD^\rho]_+]_+\bar{D}/M^3
  + {\rm h.c.}\,,
\label{def-kin}
\end{eqnarray}
where
\begin{eqnarray}
&& U_\mu = {\textstyle \frac{1}{2}}\,e^{-i\,\frac{\Phi}{2\,f}} \left(
    \partial_\mu \,e^{i\,\frac{\Phi}{f}} \right) e^{-i\,\frac{\Phi}{2\,f}} \,, \qquad \qquad 
    \Gamma_\mu ={\textstyle \frac{1}{2}}\,e^{-i\,\frac{\Phi}{2\,f}} \,\partial_\mu  \,e^{+i\,\frac{\Phi}{2\,f}}
+{\textstyle \frac{1}{2}}\, e^{+i\,\frac{\Phi}{2\,f}} \,\partial_\mu \,e^{-i\,\frac{\Phi}{2\,f}}\,,
\nonumber\\
&& \chi_\pm = {\textstyle \frac{1}{2}} \left(
e^{+i\,\frac{\Phi}{2\,f}} \,\chi_0 \,e^{+i\,\frac{\Phi}{2\,f}}
\pm e^{-i\,\frac{\Phi}{2\,f}} \,\chi_0 \,e^{-i\,\frac{\Phi}{2\,f}}
\right) \,, \qquad \chi_0 =2\,B_0\, {\rm diag} (m_u,m_d,m_s) \,,
\nonumber\\
&& \CD_\mu \bar D = \partial_\mu \, \bar D + \Gamma_\mu\,\bar D \,, \qquad \qquad \qquad \quad \;\; 
\CD_\mu D = \partial_\mu \,D  - D\,\Gamma_\mu \,.
\label{def-chi}
\end{eqnarray}
The quark masses enter via  $\chi_\pm$ and
the octet of the Goldstone bosons is encoded into the $3\times3$ matrix $\Phi$.
The covariant derivative $\CD_\mu$ in the kinetic term of the $D$ mesons generates the leading order two-body chiral interaction, recognized as the  Weinberg-Tomozawa interaction. Its interaction strength is determined by the parameter $f$, the chiral limit value of the pion-decay constant. The parameter $M$ measures the mass of the $D$ mesons in the chiral limit, provided that a suitable renormalization scheme is applied \cite{Lutz:2018cqo,Guo:2018kno}. We consider the isospin limit with $m_u = m_d = m$. The hadronic decay width of the charged $D^*$-meson implies $|g_P| = 0.57 \pm 0.07 $ \cite{Lutz:2007sk}.
Further symmetry breaking counter terms involving two $\chi_+$ fields are not shown in (\ref{def-kin}) but are systematically considered in \cite{Guo:2018kno}. In the latter work it was illustrated in great detail that a chiral decomposition of the charmed meson masses is well converging if it is organized in terms of on-shell meson masses, rather than bare masses. 
This implies that the derivation of the charmed mesons masses on a given lattice ensemble requires the solution of a coupled and non-linear set of four equations. 

The role of subleading operators of chiral order $Q^3$ as introduced first in \cite{Du:2017ttu} was explored. Such terms, proportional to $g_i$ in (\ref{def-kin}), do not affect the charmed-meson masses but do impact the scattering phase shifts.

\begin{figure}[t]
\center{
\includegraphics[keepaspectratio,width=0.7\textwidth]{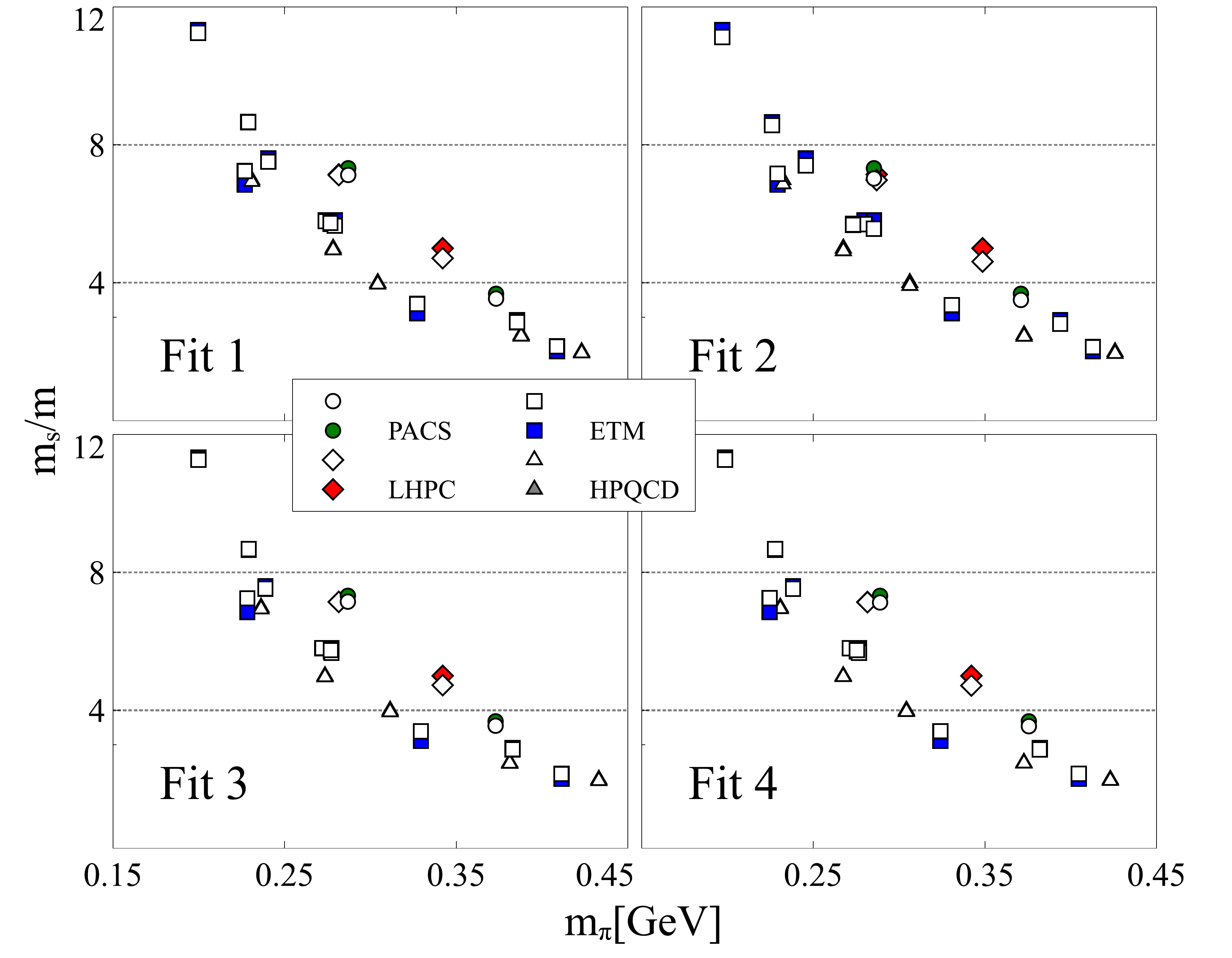} }
\vskip-0.2cm
\caption{\label{fig:ratio} 
The quark-mass ratios $m_s/m$ are shown for the various lattice ensembles considered. Closed symbols show the values from the lattice collaborations, open symbols show our results. }
\end{figure}

\section{Fit to QCD lattice data}

We determine the LECs of the chiral Lagrangian from lattice QCD simulations of the $D$-meson masses. HPQCD  provides a data set for the pseudoscalar $D$-meson masses  \cite{Na:2012iu}, based on the MILC AsqTad ensembles \cite{Bazavov:2009bb}. Those ensembles are also used in \cite{Liu:2012zya}, however employing domain-wall quarks as set up by LHPC. This work provides the pseudoscalar charmed meson masses but also some s-wave scattering lengths, that  characterize  the low-energy interaction of the charmed mesons with the Goldstone bosons. Based on the PACS-CS ensembles, the masses for charmed mesons with both $J^P=0^-$ and $1^-$ are calculated in \cite{Mohler:2011ke,Lang:2014yfa}. An even richer data set derived 
on the ETMC ensembles is published in \cite{Kalinowski:2015bwa}. So far charmed meson masses were
computed on one HSC ensemble only \cite{Moir:2016srx}. However,  besides the masses, this work also considers the $\pi D$ scattering process on that ensemble.

\begin{table}[t]
\setlength{\tabcolsep}{2.5mm}
\renewcommand{\arraystretch}{1.1}
\begin{center}
\begin{tabular}{l|rrrr}
                                            &  Fit 1     &  Fit 2    & Fit 3    & Fit 4      \\ \hline
$ M\;\;$ \hfill [GeV]                       &  1.8762    &  1.9382   &  1.9089  & 1.8846  \\ \hline

$ c_0$                                      &  0.2270    &  0.3457   &  0.2957  &  0.3002  \\
$ c_1$                                      &  0.6703    &  0.9076   &  0.8765  &  0.8880  \\
$ c_2$                                      & -0.6031    & -2.2299   & -1.6630  & -1.3452  \\
$ c_3$                                      &  1.2062    &  4.5768   &  3.3260  &  3.0206  \\
$ c_4$                                      &  0.3644    &  2.0012   &  1.2436  &  0.9122  \\    
$ c_5$                                      & -0.7287    & -4.1445   & -2.4873  & -2.1393  \\    \hline
$ g_1\;\;$\hfill [GeV$^{-1}$]               &  0         &  0        &  0.4276  &  0.4407  \\
$ g_2\;\;$\hfill [GeV$^{-1}$]               &  0         &  0        &  1.0318  &  0.8788  \\
$ g_3\;\;$\hfill [GeV$^{-1}$]               &  0         &  0        &  0.2772  &  0.2003  

\end{tabular}
\caption{The low-energy constants (LEC) from four fit scenarios as explained in \cite{Guo:2018kno}. Each parameter set reproduces the isospin average of the empirical D and $D^*$ meson masses from the PDG. 
The value $f = 92.4$ MeV was used in \cite{Guo:2018kno}.
}

\label{tab:1}
\end{center}
\end{table}

We perform chiral extrapolations based on the finite-box framework originally set up for chiral extrapolations of the baryon masses \cite{Lutz:2014oxa}. Discretization effects are not implemented so far. Only the data sets with the pion and kaon masses smaller than about 600 MeV are considered. The empirical $D$-meson masses from the PDG are used as constraints in our analysis defining  our non-standard lattice scale settings. The low-energy constants (LEC) are obtained by a global fit to the QCD lattice data set. Most ensembles suffer from a sizeable uncertainty in the choice of their charm quark mass, not always hitting its physical value. Therefore our fits consider in most cases  mass splittings of the charmed mesons only.
An \textit{ad hoc} systematic error for the charmed-meson masses is imposed. By requiring that the 
$\chi^2$ per data point turns close to one, we arrive at our estimate of 5-10 MeV for the latter.

For a given ensemble the quark masses, $m= (m_u+ m_d)/2$ and $m_s$, are determined from their published pion and kaon masses. This involves the low-energy parameters of Gasser and Leutwyler, for which we derived particular values in \cite{Guo:2018kno}. The physical quark-mass ratio $m_s/m$   is compatible with the latest result of ETMC \cite{Carrasco:2014cwa} with $m_s/m = 26.66(32)$ always. In Fig.\,\ref{fig:ratio}, the quark-mass ratio $m_s/m$ on various lattice ensembles are shown. All our four fit scenarios in \cite{Guo:2018kno} lead to a good agreement with the lattice results, even though such ratios did not enter any of our chisquare functions. This is an important  result, since it justifies our choices for the quark masses on the HSC ensemble. Note that such ratios are not available from HSC directly.

In Tab.\,\ref{tab:1} we recall our values for the low-energy parameters $M $, $c_i$ and $g_i$ from \cite{Guo:2018kno}. All four scenarios recover the open-charm meson masses on the various ensembles as discussed above but also  the s-wave scattering lengths from \cite{Liu:2012zya}. In addition, Fit 2-4 are adjusted to the scattering phase shifts of HSC \cite{Moir:2016srx}. In Fit 3 and 4, the subleading counterterms (\ref{def-kin}) are activated. We note that Fit 1 and 3 imposes the relations $c_2 = -c_3/2$ and $c_4 = -c_5/2$ which hold in the large $N_c$ limit   
of QCD.

\section{Phase shifts and poles in the complex plane}

In this section we discuss the coupled-channel dynamics of $J^P=0^+$ charmed meson.
We apply the on-shell reduction scheme as developed in\,\cite{Lutz:2001yb} to derive the coupled-channel unitarized scattering amplitude. This approach rests on a matching scale $\mu_M$, the natural value of which is given in \cite{Kolomeitsev:2003ac}. Given the set of low-energy parameters in  Tab. \ref{tab:1} phase shifts and inelasticity parameters are determined in all 
flavous sectors characterized by isospin ($I$) and strangeness ($S$) quantum numbers. 

\begin{table}[t]
\setlength{\tabcolsep}{2.5mm}
\renewcommand{\arraystretch}{1.5}
\begin{center}
\hspace*{+0.25cm}
\begin{tabular}{c|c|c|c}
\hline
            &$(I,S)= (1,1)$      &$(I,S)= (1/2,0)$    &$(I,S)= (0,-1)$     \\ \hline
WT          & $2.488_{-19}^{+22}-0.083_{-5}^{+14}i$  &   $2.390_{-17}^{+20}-0.038_{-1}^{+0}i$ &   $2.335_{+15}^{-43}$       \\ 
Fit 1       & $2.542_{-16}^{+15}-0.114_{-9}^{+19}i$  &   $2.471_{-7}^{+8}-0.046_{-3}^{+7}i$   &   $2.360_{-0}^{-1}-0.143_{-14}^{+17}i$ \\ 
Fit 2       & $2.450_{-9}^{+8}-0.297_{-8}^{+10}i$    &   $2.460_{-11}^{+17}-0.152_{+2}^{-5}i$ &   $2.287_{+2}^{-4}-0.124_{-12}^{+14}i$ \\ 
Fit 3       & $2.389_{-9}^{+6}-0.336_{-6}^{+11}i$    &   $2.463_{-27}^{+37}-0.106_{+6}^{-8}i$ &   $2.230_{+3}^{-4}-0.121_{-11}^{+13}i$ \\ 
Fit 4       & $2.382_{-10}^{+10}-0.322_{-10}^{+12}i$ &   $2.439_{-32}^{+42}-0.092_{+3}^{-7}i$ &   $2.229_{+3}^{-4}-0.083_{-11}^{+13}i$ \\ \hline
\end{tabular}
\caption{Pole masses of the $0^+$  meson resonances in the flavour sextet channels, in units of GeV. The $(1,1), (1/2,0), (0,-1)$  poles are located on the $(-,+), (-,-,+), (-)$ sheets respectively according to the notation used in \cite{Guo:2018gyd}. The asymmetric errors are estimated by varying the matching scale $\mu_M$ around its natural value by 0.1 GeV. With 'WT' we refer to the leading order scenario that relies on the parameter $f = 92.4$ MeV only.}
\label{tab:2}

\end{center}
\end{table}

\begin{figure}[b]
\center{
\includegraphics[keepaspectratio,width=0.38\textwidth]{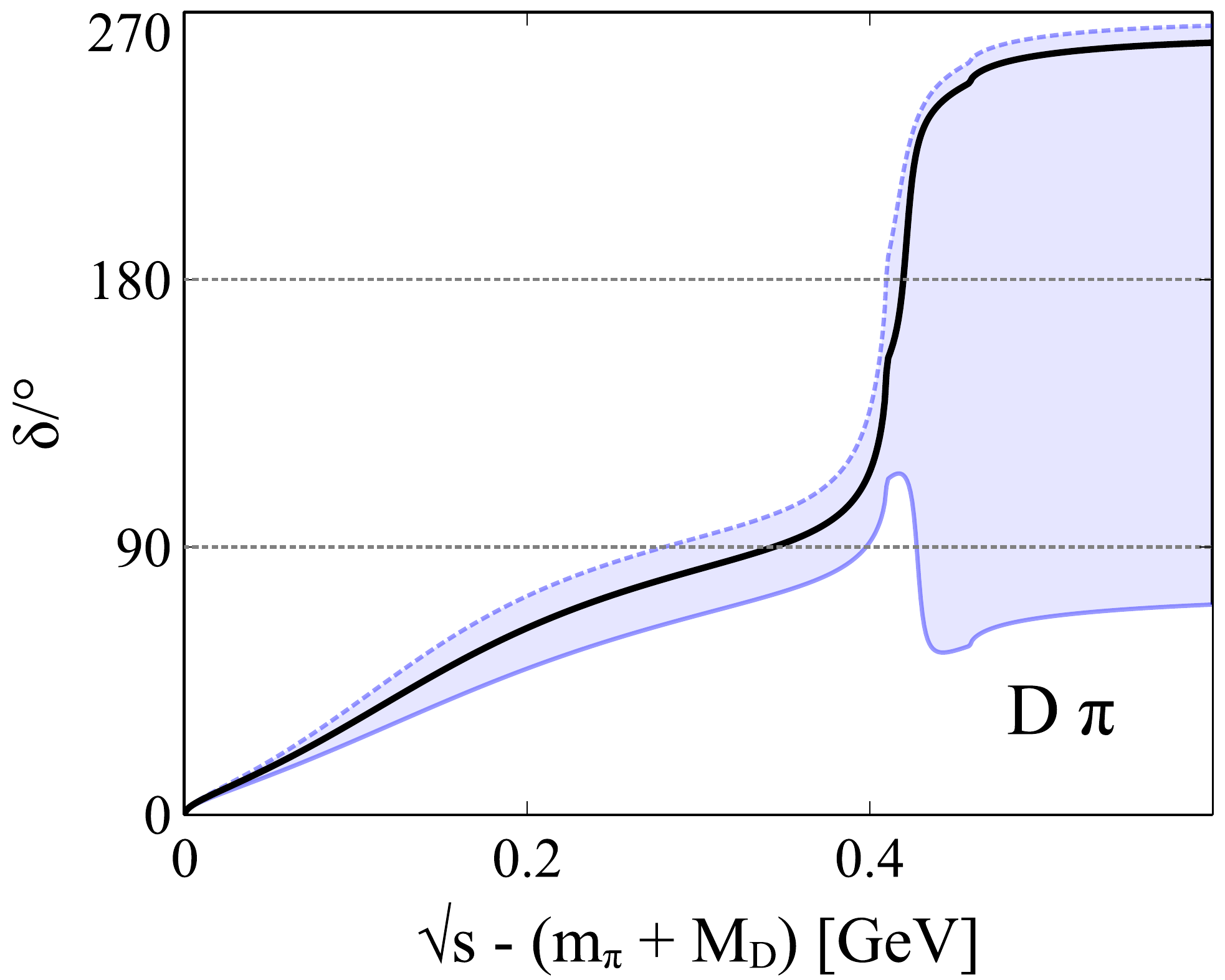}
\includegraphics[keepaspectratio,width=0.38\textwidth]{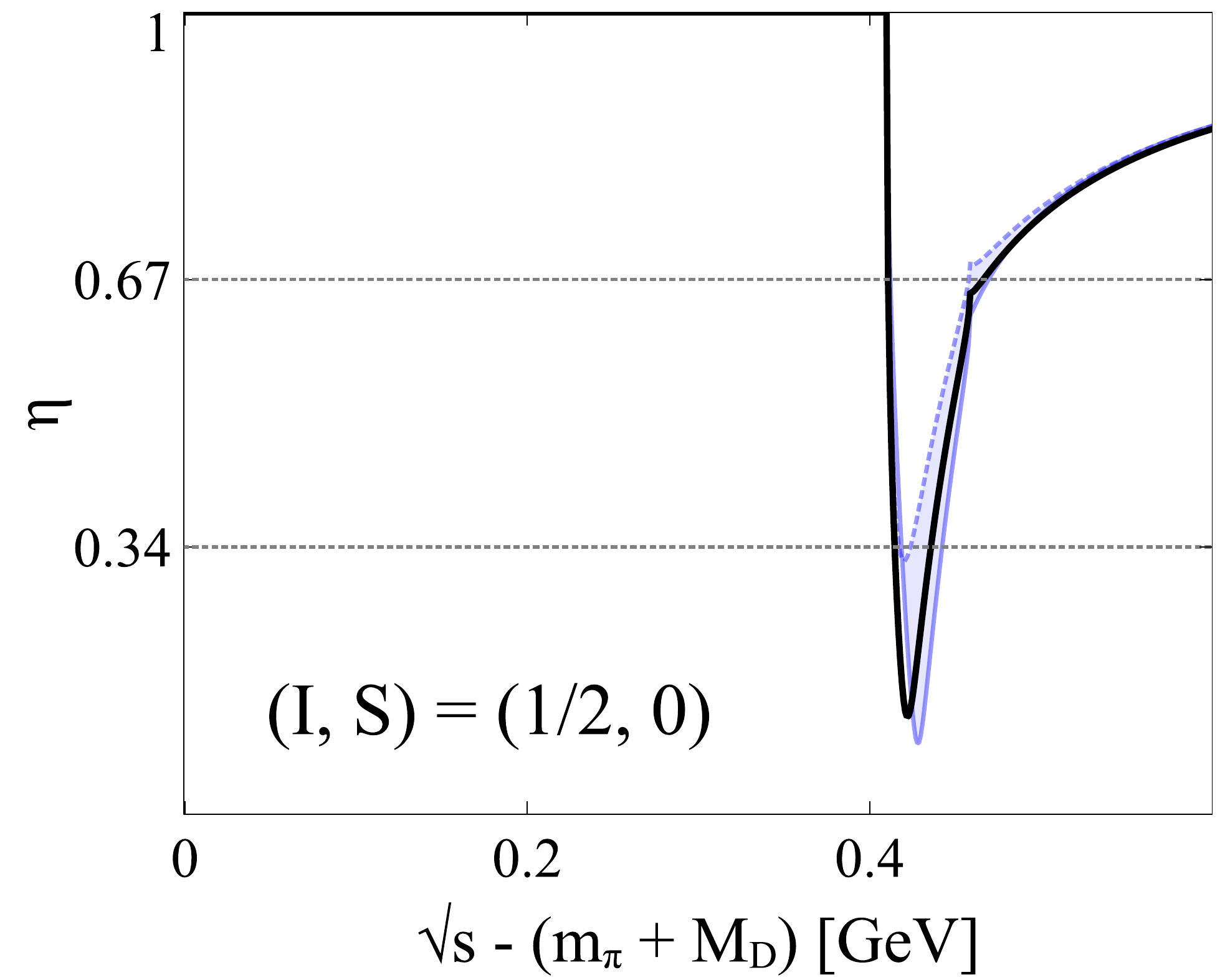}}
\vskip-0.2cm
\caption{\label{fig:phases_a} Predictions for $(I,S)=(1/2,0)$ $\pi D$ phase shifts and inelasticities from Weinberg-Tomozawa terms at physical quark masses. The blue band indicates the uncertainty by allowing $\pm 0.1$GeV deviation of $\mu_M$ from its natural value.}
\end{figure}

\begin{figure}[t]
\center{
\includegraphics[keepaspectratio,width=0.8\textwidth]{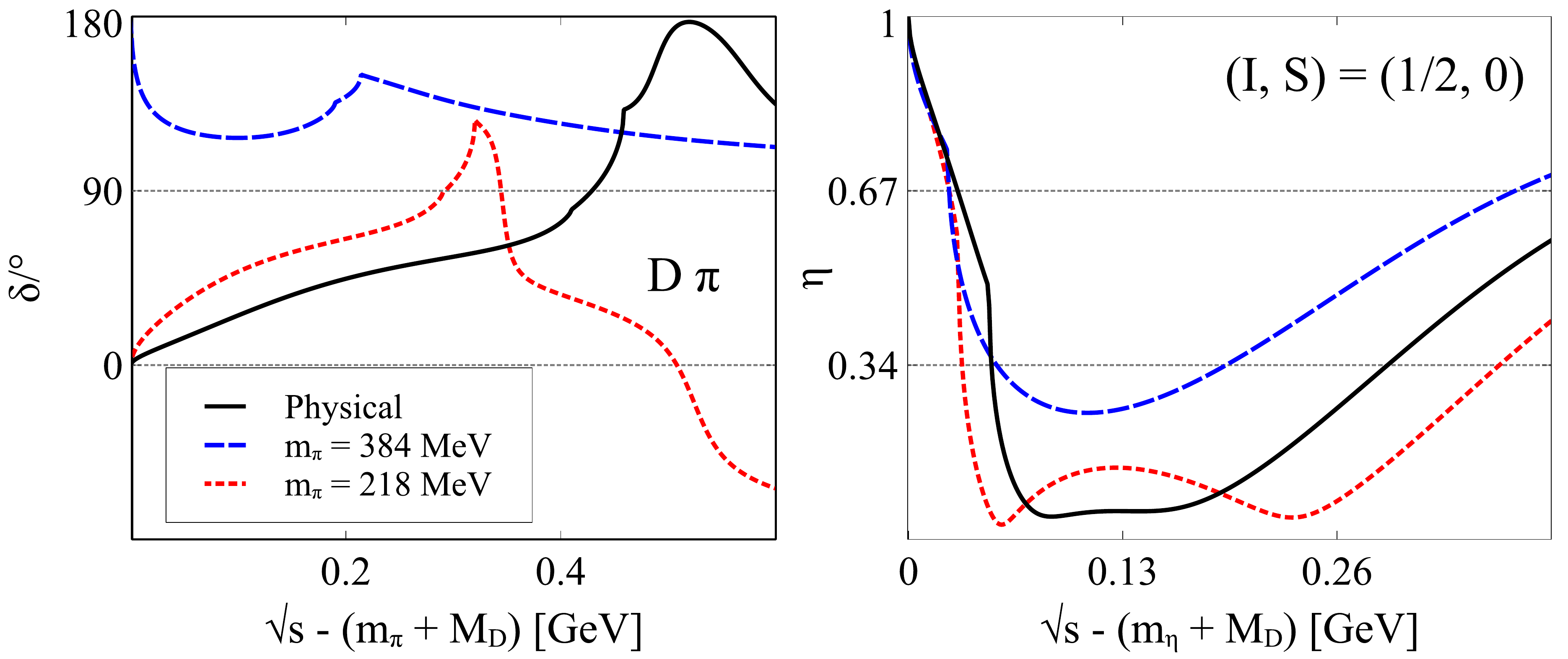}}
\vskip-0.2cm
\caption{\label{fig:phases_b} Predictions for $(I,S)=(1/2,0)$ $\pi D$ phase shifts and inelasticities from Fit 4 for the physical point but also for pion and kaon masses as given by HSC \cite{Moir:2016srx,Wilson:2015dqa}.}
\end{figure}

In this Proceeding we focus on the $\pi D$ phase shift with $(I, S) =(1/2, 0)$.
The phase shift and inelasticities generated by the leading order Weinberg-Tomozawa interaction are shown in Fig.\,\ref{fig:phases_a}. The rapid rise of the phase shift through 90$^\circ$ and 180$^\circ$ reflects the presence of a broad and a narrow resonance state in this channel. The narrow one seen around the $\eta D$ threshold is a member of the flavour sextet. The uncertainty band is implied by a variation of $\pm 0.1$ GeV  in the matching scale $\mu_M$ around its natural value \cite{Kolomeitsev:2003ac}. In comparison, the $\pi D$ phase shift  is shown in Fig.\,\ref{fig:phases_b} from our preferred Fit 4. The result at physical quark masses is shown in a  black sold line. We clearly see a signal of a resonance in between the $\eta D$ and $\bar K D_s$ thresholds. 
Most striking are our predictions for the quark-mass dependence of the $\pi D$ phase shift. We present the phase shifts at different unphysical quark masses in dashed and dotted lines in Fig.\,\ref{fig:phases_b}.

In all the sextet channels, poles are found in the lower complex plane, following the analytic continuation method illustrated in\,\cite{Guo:2018gyd}.
The pole masses are listed in Tab.\,\ref{tab:2}. 
The pole at $(I,S)=(1/2,0)$ sector is lying well between the $\eta D$ and $\bar K D_s$. The width depends on fitting scenarios, but is always significantly smaller than the antitriplet partner in the same channel. The latter exhibits a pole at $(2.082_{+2}^{-9} -0.187_{-28}^{+44}i)$\,GeV (Fit 4) where the asymmetric error is implied by a $\pm 0.1$GeV deviation of $\mu_M$ from their natural values \cite{Kolomeitsev:2003ac}.

\section{Summary}

We studied the chiral extrapolation of charmed meson masses based on the three-flavour chiral Lagrangian. About 80 lattice data points from 5 different lattice groups are analyzed.
Such data pose a significant constraint on the low-energy constants of the chiral Lagrangian 
formulated for charmed meson fields. The implication of higher order counter terms in the coupled-channel dynamics of the open-charm sector of QCD is explored. A striking quark-mass dependence of phase-shifts and inelasticity parameters is derived. At the physical point we predict a clear signal for the flavour sextet in the $\pi D$ phase shift with a pole lying in 
the complex plane between the $\eta D$ and $\bar K D_s$ thresholds.

\bibliographystyle{JHEP}
\bibliography{thesis}
\end{document}